\documentclass[10pt,twocolumn]{article}
\usepackage{ol}

\usepackage{amsmath,amssymb}
\usepackage{tabularx}
\usepackage{graphicx}

\begin{document}

\title{Fresnel Interferometric Imager: ground-based prototype}
%\title{The Fresnel Interferometric Imager: ground-based prototype}
 
\author {Denis Serre$^{1,2}$, Paul Deba$^1$, Laurent Koechlin$^1$}
\address{$^1$Laboratoire d'Astrophysique de Toulouse-Tarbes, Universit\'e de Toulouse, CNRS, 14 avenue Edouard Belin, 31400 Toulouse, France}
\address{$^2$Leiden Observatory, Leiden University, PO Box 9513, 2300RA Leiden, The Netherlands}

\maketitle

%\onecolumn%[%begin of two columns
%\begin{abstract}
\vspace*{0.5cm}
{\sl{\noindent To be published in Applied Optics Vol. 48, Iss. 15, pp. 2811-2820 (2009)}}
\section*{Abstract}
{\footnotesize{
The Fresnel Interferometric Imager is a space-based astronomical telescope project yielding milli-arc second angular resolution and high contrast images with loose manufacturing constraints. This optical concept involves diffractive focusing and formation flying: a first "primary optics" space module holds a large binary Fresnel Array, and a second "focal module" holds optical elements and focal instruments that allow for chromatic dispersion correction.\\

We have designed a reduced-size Fresnel Interferometric Imager prototype and made optical tests in our lab, in order to validate the concept for future space missions. 
The Primary module of this prototype consists of a square, 8 cm side, 23 m focal length Fresnel array. The focal module is composed of a diaphragmed small telescope used as "field lens", a small cophased diverging Fresnel Zone Lens (FZL) that cancels the dispersion and a detector. An additional module collimates the artificial targets of various shapes, sizes and dynamic ranges to be imaged. 

In this paper, we describe the experimental setup, different designs of the primary Fresnel array, and the cophased Fresnel Zone Lens that achieves rigorous chromatic correction. We give quantitative measurements of the diffraction limited performances and dynamic range on double sources. The tests have been performed in the visible domain, $\lambda$ = 400 - 700 nm.

In addition, we present computer simulations of the prototype optics based on Fresnel propagation, that corroborate the optical tests. This numerical tool has been used to simulate the large aperture Fresnel arrays that could be sent to space with diameters of 3 to 30 m, foreseen to operate from Lyman $\alpha$ (121 nm) to mid I.R. (25$\mu$m).
}}
%\end{abstract}
%\twocolumn
%]%end of twocolumns

%%%%%%%%%%%%%%%%%%%%%%%%
\section{Introduction}

The Fresnel Interferometric Imager is a space-based telescope concept providing high angular resolution images, and on sparse fields very high dynamic range.  Its operational range spans the U.V, visible and I.R domains, from typ. 100 nm to 25 $\mu$m. This telescope uses no reflective nor refractive devices (no mirrors, no lenses) as entrance pupil, but instead an interferometric array, involving hundreds thousands of "basic" subapertures, i.e., mere holes punched in a large and thin opaque foil. Their positioning law, which is close to that of a Soret (or Fresnel) Zone Plate (FZP), causes focalisation by diffraction and interference.\\

Using Fresnel Zone Plates for high angular resolution imaging in astronomy is not in itself a novel idea. Already Baez in 1960 and 1961 \cite{Baez_nature_1960,Baez_josa_1961} proposed the use of FZP especially for UV and X-ray imaging, and since the 1990's many authors have assessed their potential for visible and infrared imaging, e.g., Chesnokov in 1993 \cite{Chesnokov_sb_1993}, Hyde in 1999 \cite{Hyde_ao_1999}, Massonnet in 2003 \cite{Massonnet_brevet_cnes}. One of the limitations of the concept usually considered is the narrow spectral bandwidth due to the high dependance of the focal length with the wavelength.\\

\begin{figure}[!htbp]
\centerline{\includegraphics[width=0.85\linewidth]{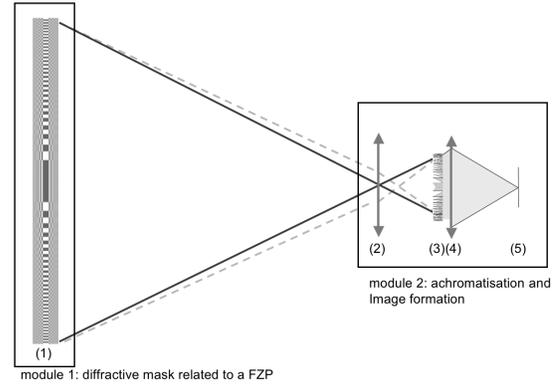}}
\caption[Achromatisation principle]
{\footnotesize In this two-module configuration, a binary diffracting mask (Fresnel array), related to a Fresnel Zone Plate, is placed on plane (1) and is used at its first order of interference. From a source at infinity, different wavelengths (dashed and dotted lines) are focussed at different distances.\newline
In the second module, field optics (2) form a pupil plane (3), where a diverging cophased Fresnel Zone Lens (FZL) is placed. Theoretically, the emerging beam is perfectly achromatic (Faklis \& Morris 1989 \cite{Faklis_oe_1989}), but divergent. A lens (4) is placed to make it converge. The final achromatic image is formed onto plane (5).}
\label{fig:achromatisation_scheme}
\end{figure}

In the concept presented in Koechlin et al 2005 \cite{Koechlin_aa_2005} and in this article, the improvements are directed toward two points:\\
1. the use of an orthogonal geometry for the FZP and correlatively that of the vacuum for its 'void' elements instead of a transparent material provides a very high quality wavefront, e.g., typically $\lambda / 100$, with strongly relaxed manufacturing and positioning constraints compared to interferometers or solid aperture devices;\\
2. the spectral bandwidth problem is adressed using a complimentar optical device in a focal module, forming an achromatic image onto a final focal plane (Fig. \ref{fig:achromatisation_scheme}, following Faklis \& Morris 1989 \cite{Faklis_oe_1989} and Hyde 1999 \cite{Hyde_ao_1999}).\\
% : the width of the individual zones of the Fresnel Zone Plate is conserved, but their disposition follows an orthogonal extension instead of a circular one, resulting in the possibility to use vaccum instead of a transparent support for the 'void' elements.

%Using Fresnel Zone Plates for high angular resolution imaging in astronomy is not in itself a novel idea: Baez in 1960 and 1961 \cite{Baez_nature_1960,Baez_josa_1961} already proposed the use of FZP especially for UV and X-ray imaging, and since the 1990's many authors have assessed their potentiality for visible, infrared and sub-mm imaging: Chesnokov in 1993 \cite{Chesnokov_sb_1993}, Hyde in 1999 \cite{Hyde_ao_1999}, Massonnet in 2003 \cite{Massonnet_brevet_cnes}. In the concept presented in Koechlin et al 2005 \cite{Koechlin_aa_2005} and in this article, the use of an orthogonal geometry for the FZP and correlatively that of the vacuum for its 'void' elements instead of a transparent material provides a very high quality wavefront, e.g., typically $\lambda / 100$, with strongly relaxed manufacturing and positioning constraints compared to interferometers or solid aperture devices. However, as the entrance pupil can be seen as a special kind of diffraction grating, on the one hand the efficiency is limited to a few \% , and on the other hand a complimentary optical device in a focal module is necessary to form an achromatic image onto a final focal plane (Fig. \ref{fig:achromatisation_scheme}, following Faklis \& Morris 1989 \cite{Faklis_oe_1989}). 

A previous paper (Koechlin et al 2005 \cite{Koechlin_aa_2005}) presents the optical principle, manufacturing tolerances and exoplanet detection capabilities of a space-based Fresnel Interferometric Imager. Another previous one (Koechlin et al 2008 \cite{Koechlin_expa_2008}) presents the potential astrophysical targets and the sensibilities required for different astrophysical targets. As a prerequisite for sending an innovative kind of instrument into space is thorough validation, during the last two years and thanks to a CNES\footnote{Centre National d'Etudes Spatiales} funding, we have built and tested a ground-based prototype equipped with the elements constituting a space-borne Fresnel Imager. In a recent paper (Serre, Koechlin, Deba 2007 \cite{Serre_spie_2007}), we have published the first qualitative results of this 8 cm aperture prototype.\\

In the first part of this paper, we describe the elements constituting the prototype: an improved transmission binary Fresnel array, and the design of the chromatic corrector with a diverging Fresnel Zone Lens, which is a small but essential element in the focal optics. In the second part, we present measurements of the optical performances: achromatisation efficiency, angular resolution (actually diffraction limited) and dynamic range ($\simeq10^{-6}$ for this 8 cm prototype).\\%, and the implementation of a coronographic stage that yields high dynamic range images of double sources: $\simeq10^{-6}$ for this 8 cm prototype.\\
% phrase suivante modifiee

We compare them to numerical simulations based on Fresnel propagation. These numerical simulations can be used for the much larger Fresnel arrays that would be used in a full fledged space mission, predicting a $10^{-7}$, or better, dynamic range. These simulations are just approached in this paper, as another publication will be dedicated to them in more detail.

%%%%%%%%%%%%%%%%%%%%%%%%%%%%%%%%%%
\section{Prototype design: Fresnel Array module}

This Fresnel array is a combination of opaque and void (transparent) elements. Starting with the description of the 1-Dimensional case where $x$ is the position of a point within the array, the transmission function is either $g(x)=0$ (opaque) or  $g(x)=1$ (void). At the center, $x=0$ and the optical path to the focus is $f$. From a point at position $x \ne 0$, the optical path to the focus is $\sqrt {x^2+f^2}$ and the Optical Path Difference OPD compared to the point at $x=0$ is:
%%%
\begin{equation}
\label{eq:opd}
%OPD(x)\, =  \sqrt {x^2+f^2} -f \;\;\; ; \;\;\; 
OPD(x)\, =  \sqrt {x^2+f^2} -f 
%k_\mathbb{R}(x)\, =  \sqrt {x^2+f^2} -f \;\;\; ; \;\;\;  k_\mathbb{R}(x) \in \mathbb{R^+}
\end{equation}
%%%
Assuming $OPD(x) = \lambda \,  k_\mathbb{R}(x)$ with $\lambda$ an arbitrary wavelength and $k_\mathbb{R}(x) \in \mathbb{R^+}$, the transmission function  $g(x, \epsilon)$ is constructed as follows:\\
\newline
$g(x, \epsilon)=0$ if $k_\mathbb{R}(x)$ modulo[1] $\in [-\alpha - \epsilon \; ; \;0.5 - \alpha + \epsilon \;[$, \\
$g(x, \epsilon)=1$ if $k_\mathbb{R}(x)$ modulo[1] $\in [0.5-\alpha + \epsilon \; ; 1- \alpha - \epsilon\;[$,\\
\newline
$\alpha \in [0,1[$ is related to a constant phase offset, and $0 \le \epsilon \ll 1$ induces a slight increase in the size of the opaque elements versus that of the void ones. 
%$\alpha$ being related to a constant phase offset $(\alpha \in [0,1[\;)$, and $\epsilon$ to a slight increase in size of the opaque versus void elements. 
$g(x, \epsilon)$ is a pseudoperiodic function with a period corresponding to the width of a Fresnel zone.

To construct a two dimensional Fresnel array, we have designed two types of geometries: a pure orthogonal one presented in Sect.\ref{sec:orth_dvpment}, and a radial-based one presented in Sect.\ref{sec:circ_dvpment}. 
%%%
\subsection{Orthogonal development  $g(x) \rightarrow T(x,y)$ \label{sec:orth_dvpment}}
\begin{figure}[!htbp]
\begin{center}
\includegraphics[width=0.9\linewidth]{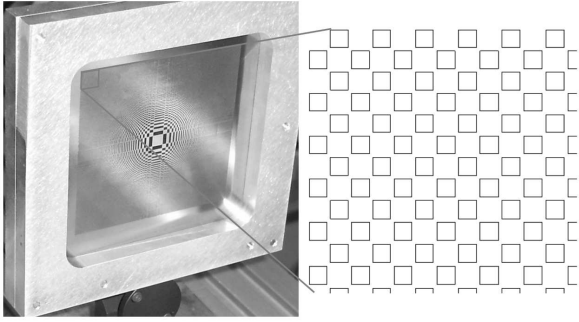}
\end{center}
\caption[Carved orthogonal Fresnel Array]
{\footnotesize Orthogonal Fresnel array, 8 cm side to side, 116 Fresnel zones in the diagonal direction (26680 subapertures) yielding a 23 m focal length at 600 nm, carved by a UV laser beam in a 80$\mu$m thick stainless steal foil. The smallest patterns (located on the periphery of the array) are 140$\mu$m squares. On the right part of the figure can be seen a zoom on these patterns, on the Autocad file generated to carve the array.}
\label{fig:grille_orth}
\end{figure}

This geometry has been used since october 2005 (see Fig. \ref{fig:grille_orth}) and is presented in ref. Koechlin et al 2005 \cite{Koechlin_aa_2005}. % and Serre et al 2006 \cite{Serre_jithd_2006}. \\
Defining  $ h(y, \epsilon) = 1-g(x, -\epsilon)$, the two-dimensional transmission function can be constructed as:
%%%
\begin{equation}
\label{eq:orth_dir}
T_o(x,y) = h(x)g(y) + g(x)h(y)
\end{equation}
or its complementary
\begin{equation}
\label{eq:orth_compl}
T_c(x,y) = h(x)h(y) + g(x)g(y)
\end{equation}
%%%
This orthogonal layout has three main consequences. First, the mechanical cohesion (assming $\epsilon >0$) allows the use of vacuum instead of transparent material for the transmissive zones. Second, the aperture edges are all in the same two directions. Last but not least, the light from an incident plane wave is split by diffraction into different wavefronts: convergent, plane or divergent.

Our Fresnel Imager uses the wavefront issued from diffraction order $+1$. Seen from the focus of this wavefront, there is a $+2\pi$ phase shift from one subaperture to the next, as $k_\mathbb{R}(x)$ increments by 1. The different wavefront elements emerging from the subapertures are in phase and interfere constructively to form a compact point spread function (PSF). Outside the PSF, the field is very dark and the residual scattered light is  confined into two orthogonal spikes. Defining the {\it{efficiency}} as the ratio between the energy in the the central peak of the PSF and the quantity of energy falling on the array, this {\it{efficiency}} is $\simeq 4 \%$(Koechlin et al 2005 \cite{Koechlin_aa_2005}).\\ %: preliminary results have been presented in Serre et al 2007\cite{Serre_spie_2007}. 
The dynamic range can further be improved  by apodization: modulating $\epsilon$ as a function of $x, y$ in the Fresnel array, or applying a transmission modulation in a pupil plane downstream, as in the Apodized Square Aperture technique (Nisenson \& Papaliolios 2001 \cite{Nisenson_apj_2001}). A Phased Induced Amplitude Apodization (PIAA) scheme could also be used (Guyon 2003 \cite{Guyon_aa_2003}): either by the remapping of the intensity distribution in the pupil, or (at least partly) by the application of a shift to the centers of the subapertures, this shift being applied within the plane of the array. In this case, the amplitude of this shift will depend on the distance of the individual subapertures to the center of the array, leading to a mean phase of the emerging wavefront varying from the center to the edges: a result similar to that of the first mirror in a PIAA system.
%%%
\subsection{Radial development $g(x, \epsilon) \rightarrow g(r, \epsilon)$ \label{sec:circ_dvpment}}
\begin{figure}[!htbp]
\begin{minipage}[c]{0.49\linewidth}
\begin{center}
\includegraphics[width=0.9\linewidth]{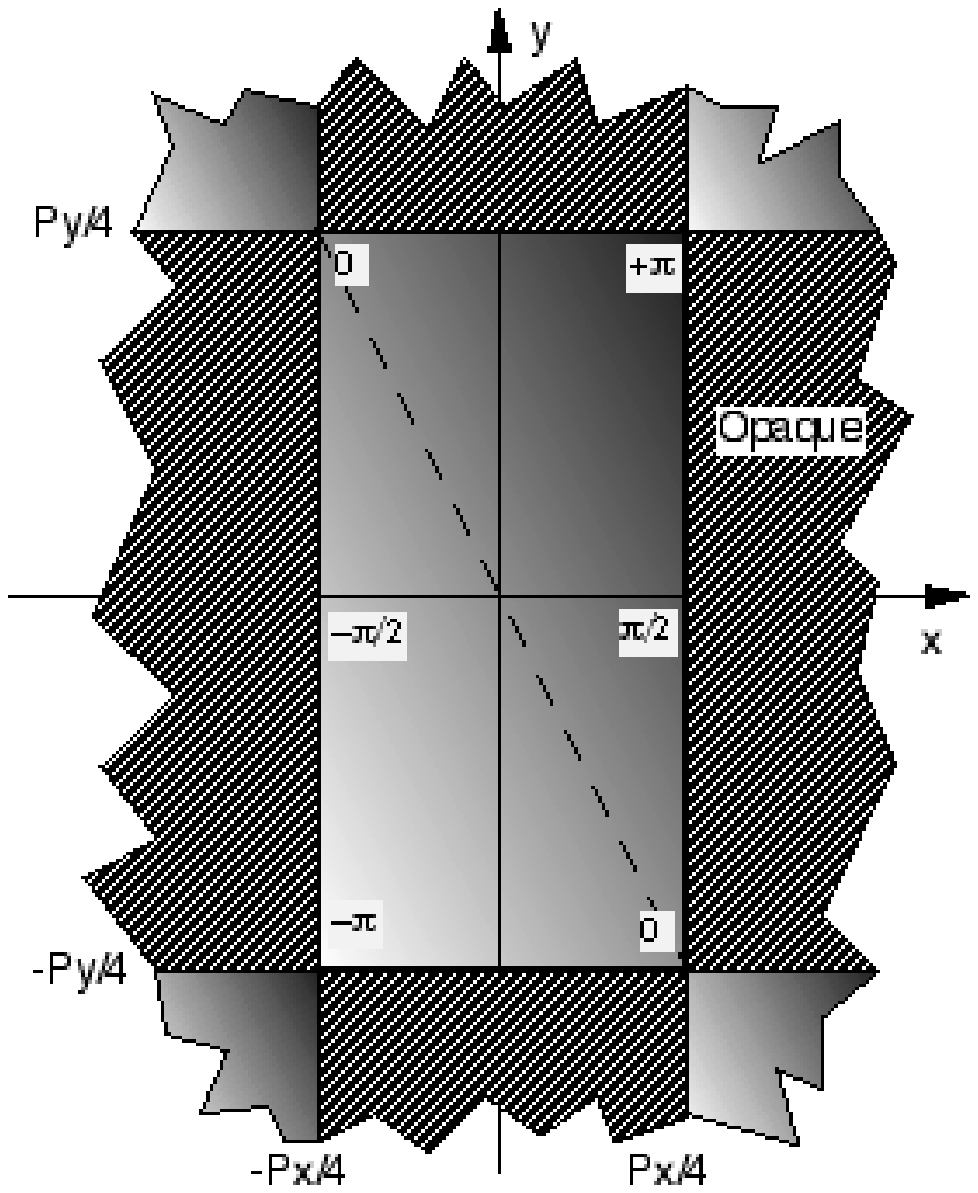}
\end{center}
\end{minipage}
\hspace*{0.5cm}
\begin{minipage}[c]{0.43\linewidth}
\begin{center}
\includegraphics[width=0.9\linewidth]{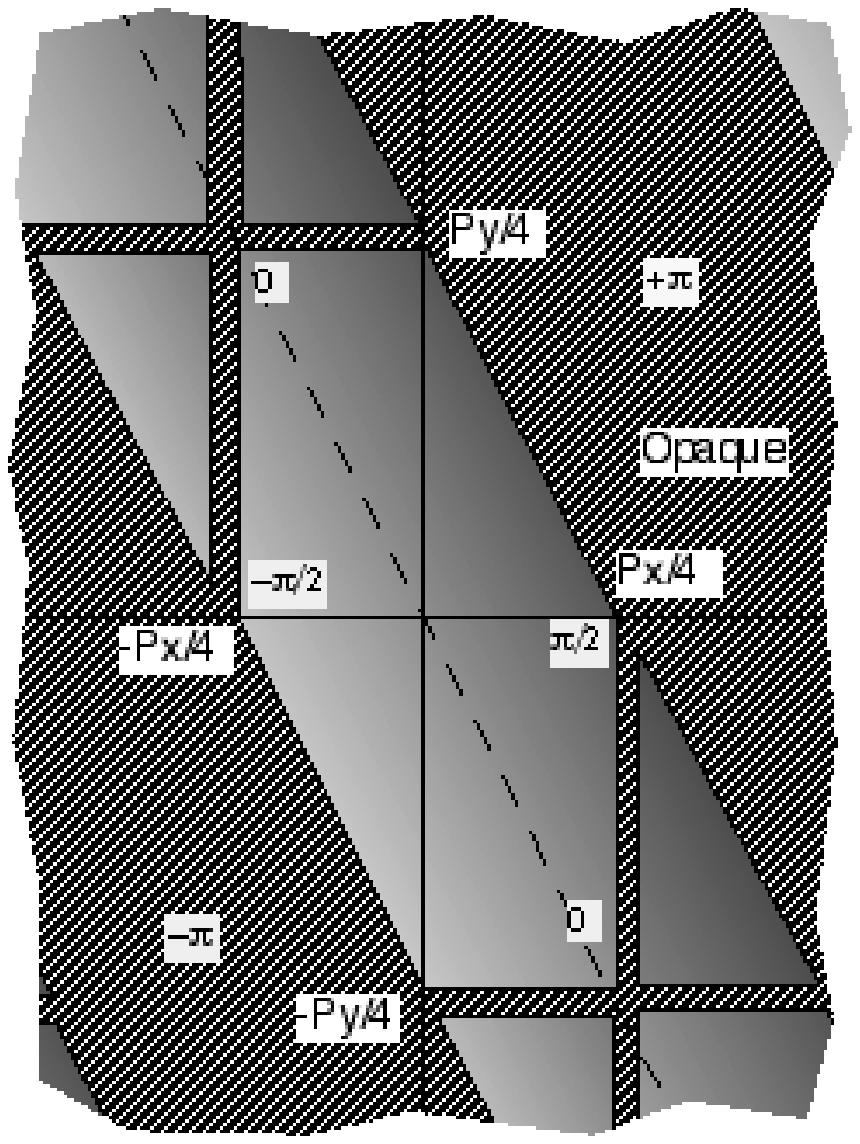}
\end{center}
\end{minipage}
\caption[Phase within a subaperture of a FZP in the orthogonal and circular cases]
{\footnotesize Left: phase map as seen from the focus in a void element in the orthogonal development of a Fresnel array. Right: phase map as seen from the focus in a void element in the multispider circular development of a Fresnel array. P$_x$ and P$_y$ are the $x$ and $y$ pseudoperiod of the local Fresnel zone.}
\label{figure_ortho}
\end{figure}

Using a radial development, a classical binary Fresnel zone plate arises (Soret 1875 \cite{Soret_1875}), leading to a nominal efficiency of 10\%. %(
%Soret ?
%%
% Waldman ? 
 %%
% Arsenault ?
 %%
 %)
 To maintain spectral span and high wavefront quality while keeping tolerances relaxed, the transparent material which could sustain the concentric rings has to be replaced by vacuum. The rings can be maintained while affecting the dynamic range as little as possible by the use of a "multispider" (see below). 
 % modif
 An alternative solution could be the use of a "photon sieve" design (Kipp 2001 \cite{Kipp_nature_2001}), but photon sieves have a low efficiency, whereas our design yields a much higher percentage of light at the focus. In addition, in our case the physical size of the underlying zones is not a problem. \\
 %end modif
The "multispider" is constituted of bars following, in each of the two orthogonal directions, a positioning law proportional to that of a 1-dimensional Fresnel zone plate. Mathematically, the thicknesses and positions of these bars follow an orthogonal development of $g(x, \epsilon_{ms})$. $\epsilon_{ms}$ is function of $x$ and negative in order to have bars thinner than the underlying Fresnel zone, and the phase shift $\alpha_{ms}$ is independent of that of the Fresnel zone plate. The effect caused by the multispider on the global transmission of the array can be minimized by adjusting its pseudoperiod to that of the Fresnel zone plate, therefore contributing to focus light.\\
\newline
As the transmissive zones are completely confined to the $[-\pi/2 ; +\pi/2]$ phase interval which was not the case for the pure orthogonal development (Fig. \ref{figure_ortho}), the transmission rate at the first order of diffraction is higher: 60\% improvement in transmission over that of an orthogonal zone plate (as the bars have to be of non-negligible thickness, the efficiency cannot reach that of a pure Soret zone plate).\\
The multispider also causes four orthogonal spikes, emaning from the same position but fainter than in the case of a pure orthogonal array. Reducing the individual bars from center to limb apodizes the multispider and reduces these brightness spikes.
% the light scattered by the multispider layout creates four spikes and slightly reduces the amount of light in the PSF, 
%If the pseudoperiod is adjusted to that of the Fresnel zone plate, the "multispider" contributes to focusing.

%\newpage
\subsection{Fresnel arrays built}

\begin{figure}[!htbp]
\begin{center}
\includegraphics[width=0.9\linewidth]{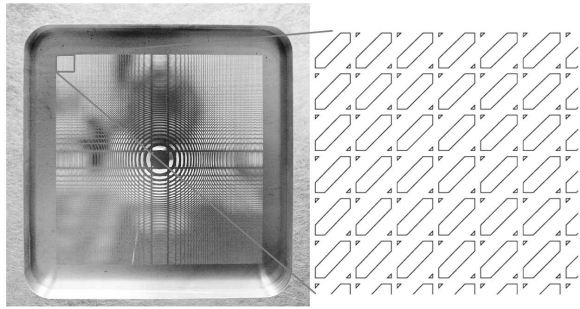}
\end{center}
\caption[Carved circular Fresnel Array with 'Multispider']
{\footnotesize Optimized Fresnel array. One can see the source collimator on the background, behind and through the array. The size, number of Fresnel zones and focal length are same as in Fig. \ref{fig:grille_orth}, but a Soret pattern held with a multispider enhances transmission by a factor 1.6 and dynamic range by a factor 2 compared to the purely orthogonal design. On the right can be seen a zoom on the patterns closest to the periphery of the array (Autocad file generated), which can be compared to the orthogonal case in Fig. \ref{fig:grille_orth}.}
\label{fig:grille_circ}
\end{figure}

The prototypes of orthogonal and radial arrays are $8 \sqrt 2$ cm in  
diagonal and 116 Fresnel zones, yielding a 23m focal length at 600nm for the order $+1$ focus. They are carved with a UV laser machine tool into a 80 $\mu$m thick metal foil (see Figs.\ref{fig:grille_orth} and \ref{fig:grille_circ}).%The numerical commands are based on eqs. \ref{eq:opd}, \ref{eq:orth_dir}, and \ref{eq:orth_compl}.

%%%%%%%%%%%%%%%%%%%%%%%%
\section{Prototype design: focal module}

\subsection{Principle and components}
A Fresnel array is very chromatic, as are Fresnel zone plates. Its focal length $F$ is wavelength dependant with:
$ F = {D^2 / 8 N \lambda}$,
 where $D$ is the diameter (or diagonal in case of a square aperture) and $N$ is the number of Fresnel zones from center to edge (corner in case of a square aperture).

\begin{figure}[!htbp]
\begin{center}
\begin{tabular}{c}
\includegraphics[width=0.95\linewidth]{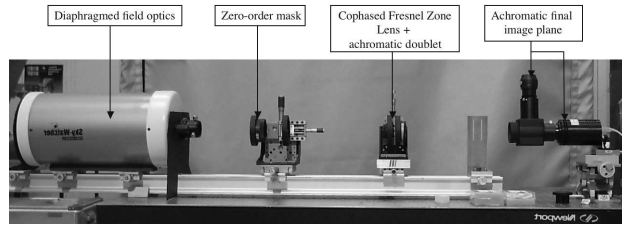}
\end{tabular}
\end{center}
\caption[Focal module of the prototype] 
{\footnotesize The focal module is composed of a Maksutov telescope used as a field lens, a cophased diverging Fresnel diffractive lens situated in the pupil plane, and an achromatic doublet next to that pupil plane. The Maksutov telescope is diaphragmed to a 3.1 cm diameter, resulting in $\Delta \lambda / \lambda = 0.4$ for a non-diaphragmed 200 arcsec field of view (Eq.  \ref{eq:BP_square}). The achromatic image plane can be sent either onto a CCD or an eyepiece for control. A small mask is placed at the focal plane of the "field telescope", eliminating residual light from the 0 and -1 diffraction orders of the Fresnel array located 23 m upstream.}
\label{fig:focal_module}
\end{figure}

The focal module (Fig. \ref{fig:focal_module}) features an application of the achromatisation scheme proposed by Schuppman in 1899 \cite{Schupmann_1899} (Fig. \ref{fig:achromatisation_scheme}): a field lens (in our case a diaphragmed two-mirror Cassegrain-Maksutov combination) produces a pupil plane, where a cophased diverging FZL is placed, the Fresnel zones of which are being superposed to the imaged Fresnel zones of the primary array. The combination of the order $+1$ of the Fresnel array and the order $-1$ of the FZL adds a $-2 \pi$ phase shift at the places where $+2\pi$ phase shifts have been created at Fresnel zone boundaries. The wavefront is restored to its original continuity and smoothness, completely wavelength independent, and produces an achromatic diverging beam (Faklis \& Morris 1989 \cite{Faklis_oe_1989}), then made convergent by an achromatic doublet downstream (Sec. \ref{sec:achromatisation_principle}).

\subsubsection{Principle of achromatisation \label{sec:achromatisation_principle}}

\begin{figure}[!htbp]
   \begin{center}
\begin{tabular}{c}
      \includegraphics[width=0.9\linewidth]{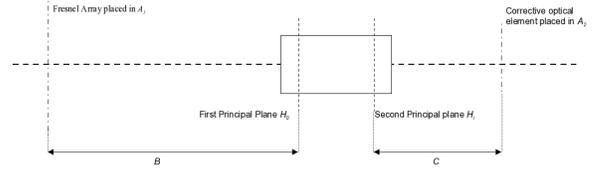}
\end{tabular}
   \end{center}
   \caption[Schupmann principle applied to the Fresnel Imager]
{\footnotesize The optical element placed in $A_2$ (power $P_2$) will correct the chromatism induced by the Fresnel array placed in $A_1$ (power $P_1$), using an optical device characterised by its principal planes $H_o$ and $H_i$ (power $P_{H_{o}H_{i}}$).}
\label{fig:demo_schupmann}
\end{figure}

Our system is an example where chromatic aberration is actually corrected (i.e., cancelled) by combining two diffractive lenses. An achromatic mirror combination acting as "field lens" conjugates the two diffractive lenses. The demonstration of the achromatisation principle can be done using ray transfer matrix analysis as in Hyde 1999 \cite{Hyde_ao_1999}. Here we present an equivalent demonstration, with the exception that we consider the principal planes of the field optics and demonstrate that not only the $-1$ order of the FZL can be used, but other orders as well. Let a Fresnel Array be placed in plane $A_1$, an optical device be representated by its principal planes $H_0$ and $H_i$, and the corrective optical element be placed in plane $A_2$; let $B$ and $C$ be the distances $A_1H_0$ and $H_iA_2$ (Fig. \ref{fig:demo_schupmann}). The purpose of the problem is to get constraints on the focal distance and size of the optical element placed in $A_2$ and the distances between the three optical elements. The ray transfer matrix can be written:\\
\newline
\hspace*{-0.3cm}$[T] = \begin{bmatrix}
T_{1,1} & T_{1,2} \\
T_{2,1} & T_{2,2}
\end{bmatrix}$

%\centerline{$[T] = \begin{bmatrix}
%T_{1,1} & T_{1,2} \\
%T_{2,1} & T_{2,2}
%\end{bmatrix}$}

\begin{equation}
\centerline{ $=
\begin{bmatrix}
1 & 0 \\
-P_{2} & 1
\end{bmatrix}
\begin{bmatrix}
1 & C \\
0 & 1
\end{bmatrix}
\begin{bmatrix}
1 & 0 \\
-P_{H_{o}H_{i}} & 1
\end{bmatrix}
\begin{bmatrix}
1 & B \\
0 & 1
\end{bmatrix}
\begin{bmatrix}
1 & 0 \\
-P_{1} & 1
\end{bmatrix}
$
}
\label{eq:transfert_total}
\end{equation}

\noindent Assuming $P_1 = \alpha \lambda$ and $P_2 = \beta \lambda$ we get:\\
\newline
\centerline{$T_{1,1} =  1 - C\,P_{H_oH_i}  - (B+C)\,\alpha \lambda + B \,C \,\alpha \, \lambda\,P_{H_oH_i} $}\\
\newline
\centerline{$T_{2,2} =  1- B\,P_{H_oH_i} -(B+C)\,\beta \lambda + B \,C \,\beta\, \lambda \,P_{H_oH_i} $}
\newline
\begin{equation}
\centerline{$T_{1,2} =  B + C - P_{H_oH_i}\,B \,C  $}
\label{eq:3_termes_matrice_transfert}
\end{equation}

Optical power $P$ can be written:\\
\newline
\centerline{$P = - T_{2,1} =  P_{H_oH_i} - \lambda [ -\beta + \beta P_{H_oH_i} C - \alpha + \alpha P_{H_oH_i} B]$}
%\newline
% - \lambda^2 \alpha \beta (B+C-P_{H_oH_i} B C)$}
\begin{equation}
\hspace*{-0.8cm}\centerline{$ - \lambda^2 \alpha \beta (B+C-P_{H_oH_i} B C)$}
%\centerline{$P = - T_{2,1} =  P_{H_oH_i} - \lambda [ -\beta + \beta P_{H_oH_i} C - \alpha + \alpha P_{H_oH_i} B] - \lambda^2 \alpha \beta (B+C-P_{H_oH_i} B C)$}
\label{eq:power}
\end{equation}

A chromatic correction requires a power independent of wavelength. The term proportional to $\lambda^2$ and $\lambda$ must therefore be cancelled. Cancelling the term proportional to $\lambda^2$ implies:
\begin{equation}
\centerline{$B+C-P_{H_oH_i}\, B\, C = 0$}
\label{eq:premiere_equa_achrom}
\end{equation}
which means that $A_2$ is a plane conjugate of $A_1$, in our case a pupil plane. Another consequence of Eq. \ref{eq:premiere_equa_achrom} is to make terms $T_{1,1}$ and $T_{2,2}$ be wavelength independent, meaning that there will be no transversal or angular differential magnification with wavelength. From equation \ref{eq:premiere_equa_achrom} we deduce that:
\begin{equation}
\centerline{$\frac{B}{C} = P_{H_oH_i}B-1$ and $\frac{C}{B} = P_{H_oH_i}C-1$}
\end{equation}
The term proportional to $\lambda$ in Eq. \ref{eq:power} will be equal to 0 ($\forall P_{H_oH_i}$) only if
\begin{equation}
\centerline{$\frac{\beta}{\alpha} = -\frac{B^2}{C^2}$}
\end{equation}
Using $\lambda F = \frac{D^2}{8N}\frac{1}{m}$ ($D$ being the diameter or diagonal of the Fresnel Array, $N$ the number of Fresnel zones involved and $m$ the interference order used) and considering that the optical element placed in $A_2$ will have a similar equation, we can write:
%On peut d'autre part \'{e}crire en utilisant que $\lambda f = \frac{c^2}{8N}\frac{1}{m}$, et en supposant que l'optique plac\'{e}e en $A_2$ suivra une loi ayant la m\^{e}me forme:% la relation \ref{eq:distance_focale}:
\begin{equation}
\centerline{$P_1 = \alpha \lambda = \frac{8N_1\lambda\,m_1}{{D_1}^2}$ and $P_2 = \beta \lambda = \frac{8N_2\lambda\,m_2}{{D_2}^2}$}
\end{equation}
and we get a second achromatisation condition after equation \ref{eq:premiere_equa_achrom}:
\begin{equation}
\centerline{$\frac{B^2}{C^2} = -\frac{N_2}{N_1}\frac{{D_1}^2}{{D_2}^2}\frac{m_2}{m_1}$}
\label{eq:deuxieme_equa_achrom}
\end{equation}
Therefore, a solution to correct for the chromatism induced by the Fresnel Array placed in $A_1$ is to place in a pupil plane $A_2$ a Fresnel Zone Lens with an equal number of zones $N_2 = N_1$, used at its order -1 and whose diameter is that of the imaged Fresnel Array. Here we join a conclusion at which Faklis \& Morris \cite{Faklis_oe_1989} arrived at in the field of chromatic correction of a holographic imaging process. Another possibility would be the use of the FZL at orders $m_2=$ -2, -3... while keeping $D_2$ constant and using $N_2 = \frac{N_1}{\mid m_2 \mid}$ numbers of Fresnel zones. But, as the efficiency would be lower for $\lambda \neq \lambda_{blaze}$ (Faklis \& Morris 1995 \cite{Faklis_ao_1995}), the only reason for using it would be the limitations of manufacturing capabilities of the most external zones. In that case, a solution would be to manufacture the external zones for an order of interference $m_2=$ -2, -3... while keeping the central zones manufactured for an order $m_2=$ -1.

\subsubsection{Compromise between wavelength bandwidth and size of field of view}
The geometric correction is rigorous in the sense that the correction done by the FZL is done for all the wavelengths intercepted. But since the field optics has a finite diameter:\\
- the wavelength bandwidth will be maximum for an on-axis source;\\
- but, for a source sufficently off-axis so that its image at the nominal wavelength formed by the FZP is at the edge of the field lens, the spectral bandwidth will be close to 0 if we { \it accept }no vignetting effect. Assuming $n$ being the diameter of the field of view in number of resolution elements\footnote {A resolution element has an angular extension $\sqrt{2} \lambda / D$ for a square pupil of diagonal $D$, and $1.22 \lambda / D$ for a circular pupil of diameter $D$.} and $\lambda$ the wavelength focussed on the field optics, the compromise between the size of the field of view and the wavelength bandwidth can be written:
\begin{equation}
\frac{\Delta \lambda}{\lambda} = \frac{2\times\text{{\it{Field Optics diameter}}} - n \sqrt{2}\frac{D}{8\,N}}{D}
\label{eq:BP_square}
\end{equation}
\begin{equation}
\frac{\Delta \lambda}{\lambda} = \frac{2\times\text{{\it{Field Optics diameter}}} - n \,1.22 \, \frac{D}{8\,N}}{D}
\label{eq:BP}
\end{equation}
with $1.22\frac{D}{8N}$ the linear size of the resolution element on the field optics (or $\sqrt{2} \, \frac{D}{8N}$ in the case of a square aperture), $D$ still being the diameter (or diagonal in case of a square aperture) of the FZP and $N$ the number of Fresnel zones from center to edge (corner in case of a square aperture). For larger fields or smaller or greater wavelengths, the  chromatic correction will still be rigourous, but, as the beam will be vignetted by the field optics, the luminosity and image quality will decrease, but may still be acceptable depending on the applications. A detailed discussion of this limitation can be found in section 3 of Koechlin 2008 \cite{Koechlin_expa_2008}.

As the beam is strongly compressed from the Fresnel array to the pupil plane formed by the field optics, the blazed secondary and convergent optics can be of modest size. In a large space-based instrument, the beam compression ratio between the Fresnel Array and the correcting FZL could reach 100 or more, the limiting parameter being the manufacturing possibilities of the smallest patterns near the edge of the FZL. In our prototype, we use a FZL etched on a fused silica plate and a compression ratio of 7, because the primary array is already small.

This FZL is blazed for high efficiency; close to 100\% at the blaze wavelength -excluding reflecting effects on entrance and exit surfaces- which improves the overall transmission of the instrument. Although theoretically perfect, this correction is bandpass limited to $\Delta \lambda / \lambda \simeq 30\%$ in practice, mainly due to the blaze angle mismatch of the secondary Fresnel lens with non optimal wavelengths (see below).%, spectral efficiency of coatings and focal instrumentation optics, and spectral response of the detector (see below).

%%%%%%%%%%%%%%%%%%%%%%%%
%\subsection{Design and efficiency of the chromatic corrector}

%%%%%
\subsection{Design of the mainpiece: chromatic corrector}

%The chromatic corrector is a Fresnel Zone Lens (FZL), cophased and diverging. 

\begin{figure}[!h]
\centering
\includegraphics[width=0.9\linewidth]{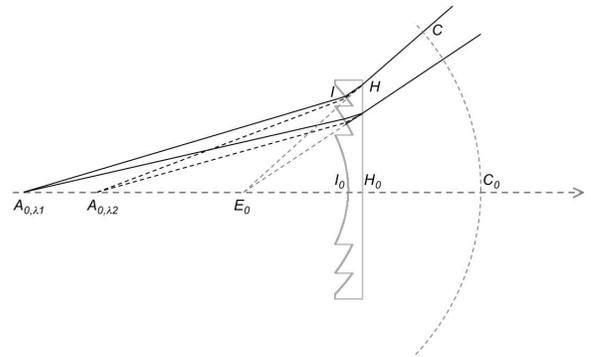}
\caption[Design of the Fresnel Zone Lens]
{\footnotesize Optical paths in a blazed diverging Fresnel zone lens (FZL) illustrated for 3 zones. At the uncorrected focus of the Fresnel array, simply reimaged by the field optics, the wavefronts at different wavelengths converge on the optical axis at different positions $A_{0,\lambda 1}$, $A_{0,\lambda 2}$, etc. The diverging FZL, having the same number of zones as the Fresnel array, is placed in the pupil plane and used at its order -1. The "reverse" chromaticity of the FZL results in all the $A_{0,\lambda}$ becoming conjugate with a unique point $E_0$, therefore achieving chromatic correction of the emerging beam. $I$ and $H$ are respectively the entrance and exit points of a wavefront sample into (out of) the Fresnel lens. $C$ is the intersection of the emerging wavefront samples with a spherical surface centered on $E_0$. Indices $0$ are for on-axis points.}
\label{fig:fig_design_FZL}
\end{figure}

%The fields optics reimages the wavelength-dependant focus formed by the primary array on the wavelength-dependant positions $A_{0, \lambda_1} $, $A_{0, \lambda_2} $... :  the role of the FZL is to replace these images by a virtual wavelength-independent image at position $E_0$ (Fig. \ref{fig:fig_design_FZL}).
The role of the FZL is to replace the images formed by the field optics and situated at different wavelength dependent positions $A_{0, \lambda_1} $, $A_{0, \lambda_2} $...  by a virtual wavelength-independent image at position $E_0$ (Fig. \ref{fig:fig_design_FZL}). The application of the chromatic correction presented in Sec. \ref{sec:achromatisation_principle} constrains a number of parameters for this lens: size, number of zones, distance between the FZL and the wavelength-dependent images re-formed by the field optics.

For a given ray, the light emerges from focus $A_{0, \lambda} $, enters the optical medium of index $n$ at point $I$ and re-emerges at point $H$. From the Fermat principle, the following relation arises between the different segments of optical path (Fig. \ref{fig:fig_design_FZL}):
%%%
\begin{equation}
A_{0,\lambda}I + k\lambda + n_{\lambda}IH - E_0H = \rm{\bf{cst_\lambda}}
\label{eq:design_FZL}
\end{equation}
$k \in \mathbb{N}$ being the index of the Fresnel zone. The jigsaw profile of the lens is given by the locus of points $I$.%, which can be obtained by solving Eq. \ref{eq:design_FZL}.

We can solve Eq. \ref{eq:design_FZL} for an optimum wavelength $\lambda_{blaze}$; the parameters that are fixed or can be obtained, are:\\
-- $A_{0,\lambda}$:  from the primary array focal length at $\lambda_{blaze}$, the field optics focal length, and the relative position between these two optical elements, the position of the primary focus at $\lambda_{blaze}$ reimaged by the field lens can be calculated, as can be the position of the pupil plane of the Fresnel imager, where the FZL is placed.\\%
%-- $A_{0,\lambda}$: the distance from $A_{0,\lambda}$ for the chosen $\lambda_{blaze}$ to $H_0$ can be calculated from the primary array focal length, the field optics focal length, and the relative position between these two optical elements: first calculation: position of the primary focus at $\lambda_{blaze}$ reimaged by the field lens, and second calculation: position of the pupil plane of the Fresnel imager, where the FZL is placed.\\%
%-- zone index $k$: as the FZL is homothetic to the Fresnel array, for a given $H$, the zone index $k$ ($k \in \mathbb{N}$) can be calculated.\\
-- zone index $k$: as the number of zones of the FZL is similar to that of the Fresnel array, for a given $H$, the zone index $k$ ($k \in \mathbb{N}$) can be calculated.\\
-- $n_\lambda$: the optical index at $\lambda_{blaze}$ of the material in which the diverging FZL is engraved.\\%: in our case, the FZL profile is computed for $\lambda_{blaze} = 600$nm, where the corresponding index for silica is $n=1.449$.\\
% typically, the position of $A_{0,\lambda}$ for which the rays cross the optical axis at the
% PAS D'ACCORD !: pkoi point nodal ? nodal image point of the field optics. \\
-- $E_0$: as the FZL must fill the pupil plane, its size is known, and its number of zones being equal to that of the primary array its focal length too: therefore, knowing the position of $A_{0,\lambda}$, the position of $E_0$ can be calculated.\\
-- $H$: the coordinates of points $H$ can be fixed by sampling the back surface (flat in our case) of the lens.\\
-- $\bf{cst_{\lambda}}$: the value of $\bf{cst_{\lambda}}$ can be calculated by the application of Eq. \ref{eq:design_FZL} on the optical axis.\\
-- The only unknown parameters remaining in Eq. \ref{eq:design_FZL} are the coordinates of point $I$. For each point $H$, the coordinates of the corresponding point $I$ are calculated, yielding as result the jigsaw profile of the FZL. This has been done numerically for a thinly sampled profile, then sent for engraving on fused silica.

\begin{figure}[!htbp]
\begin{minipage}[c]{0.49\linewidth}
\begin{center}
\includegraphics[width=0.9\linewidth]{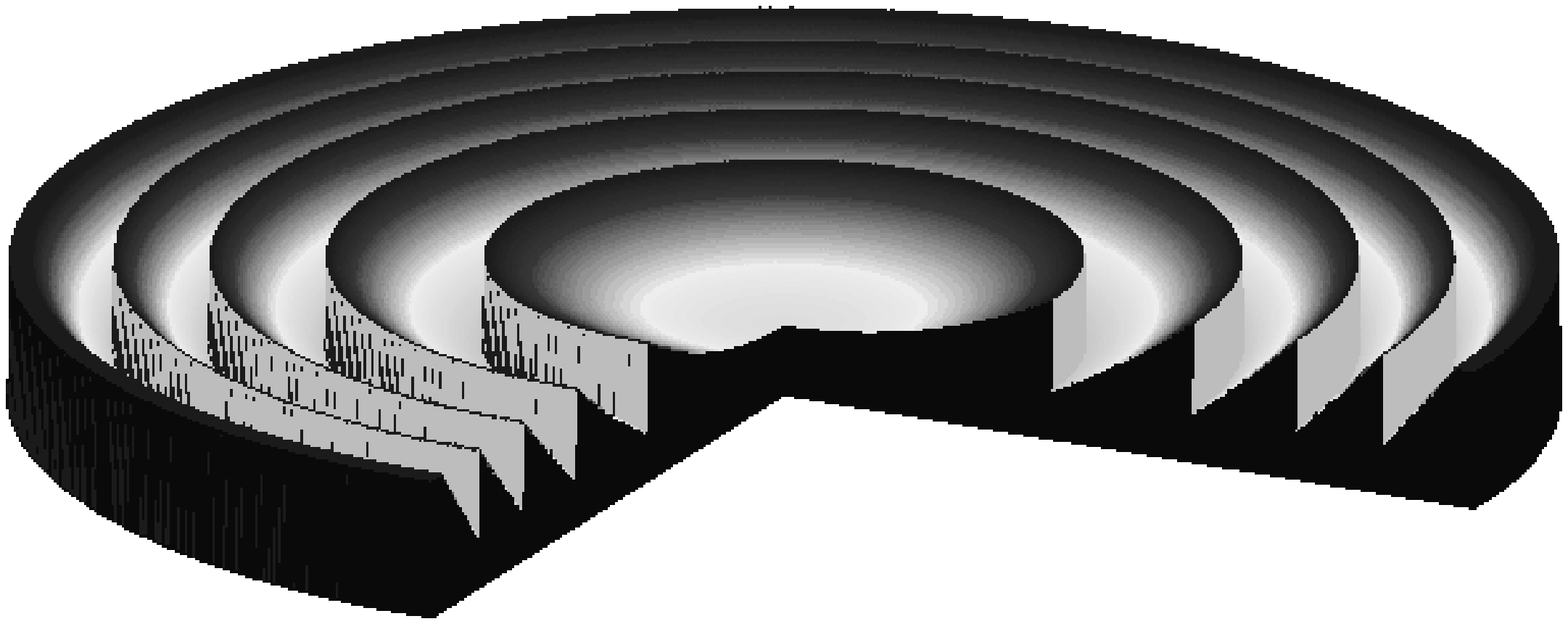}
\end{center}
\end{minipage}
\begin{minipage}[c]{0.49\linewidth}
\begin{center}
\includegraphics[width=0.9\linewidth]{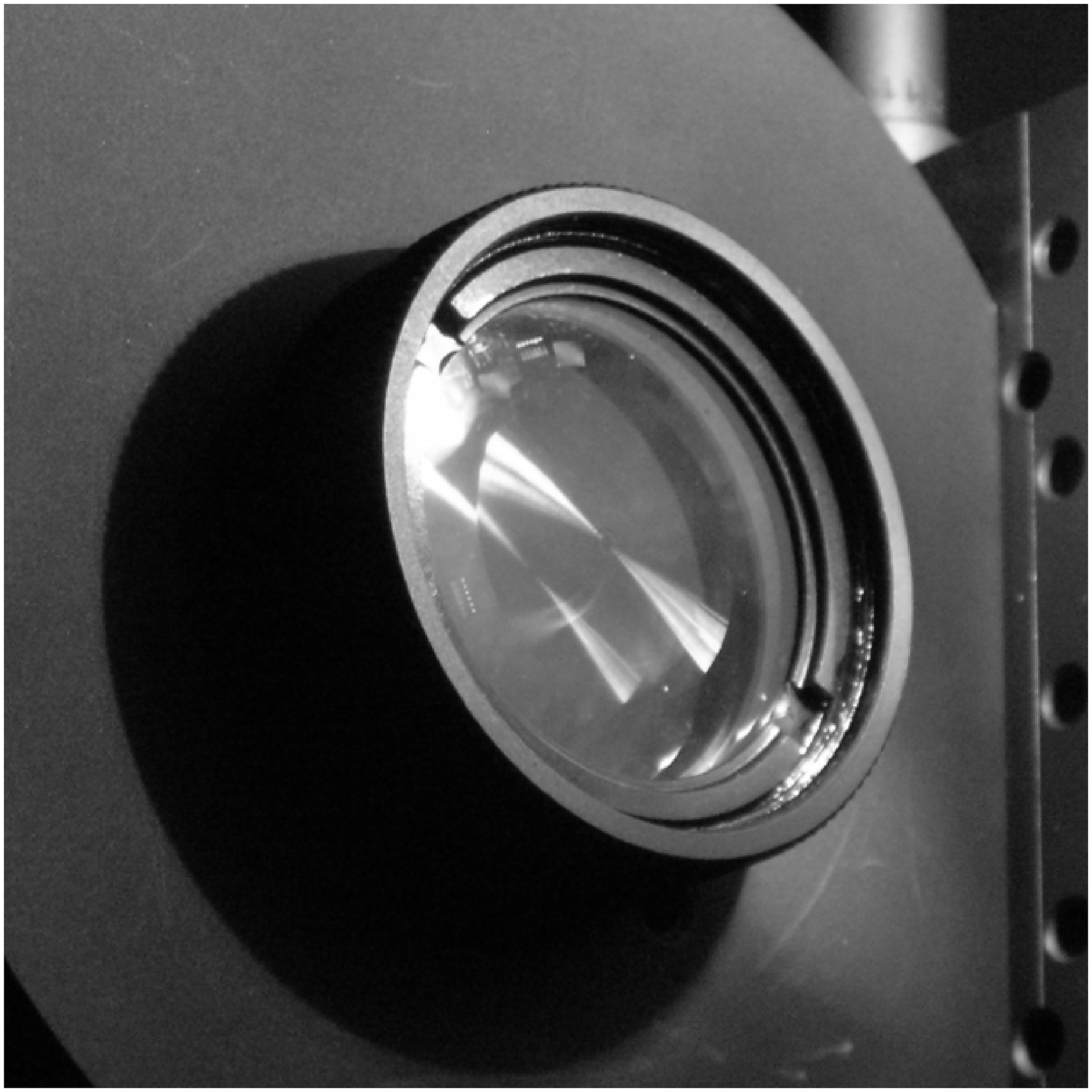}
\end{center}
\end{minipage}
\caption[Fresnel Zone Lens: computer generated view, and manufactured one]
{\footnotesize Left: 3D view of the 5 central zones of a diverging FZL. The vertical scale is highly exaggerated, as the depth of the slopes is 1.31$\mu m$ whereas the radius of the central zone is 600$\mu m$. The discretized number of levels can not be seen at the scale of this print. Right: Photography of the manufactured cophased diverging FZL: 16.054mm in effective diameter, 116 zones.}
\label{fig:FZL_realized}
\end{figure}

We have commissioned SILIOS Technologies for the realization of two 16 mm diameter and 5 mm thick FZL, blazed at 600nm with maximum depth profile 1335 nm, sampled respectively with 32 and 128 depth levels (Fig. \ref{fig:FZL_realized}). The level depth precision is $\pm $6 nm PTV, to be added to the discretization error: $\pm $6 and $\pm $22 nm respectively. The depth error is larger at the edge. With the minimal engraving width being 2 $\mu$m, and the narrowest Fresnel zone 34 $\mu$m, there can be only 16 levels within, resulting in a PTV error of  $\lambda / 16 $ for the outermost fraction of the beam.\\

As the nominal profile depth varies with $\lambda$, the FZL built are optimized only for $\lambda_{blaze} = 600$ nm. The nominal profile also varies with $n_{\lambda}$, however for $\lambda$ varying from 500 to 700nm the optical index of the fused silica varies from 1.4625 to 1.4552 (0.5\% difference), which is much less than the variation due to the wavelength itself: the height of the steps is $\frac{\lambda_{blaze}}{n-1}$, thus consequences of the dispersion due to the material are negligible.\\
Nevertheless, the profile mismatch for $\lambda \neq \lambda_{blaze}$  does not cause chromatic aberration, it causes a loss of efficiency and dynamic range for wavelengths unadapted to the blaze angle. This issue is desribed in details below.

%%%%%%%%%%%%%%%%%%%%%%%%
\section{Numerical simulations and Optical tests}

We have made numerical simulations by Fresnel propagation of the diverging FZL alone and of the complete prototype, including the primary array, field optics, zero order blocking mask, corrector blazed FZL with its discretization due to the manufacturing process, and final converging lens. 
We have also tested optically the Fresnel imager prototype on various optical sources for achromaticism, dynamic range and resolution. As the prototype is confined to a clean room, only artificial sources placed at the focal plane of a collimator have been used. 

%%%%%
\subsection{Computed Efficiency of the Fresnel diffractive lens}

Two important parameters influence the efficiency of the chromatic correction FZL: the mismatch of the wavelength used and the discretization of the profile slope levels ocurring from the manufacturing process. Several authors have studied these influences: Swanson \cite{Swanson_oe_1989} particularly studied it for the efficiency dependence with the number of levels, moreover developing the manufacturing process; Hasman \cite{Hasman_ol_1991} for the depth error consequences, and Levy \cite{Levy_josa_2001} for the analytic theory of spherical and cylindrical lenses. Faklis \& Morris \cite{Faklis_ao_1995} studied the evolution of efficiencies with wavelength (for multi-order lenses, including for first diffraction order lenses). For any wavelength, Eq. \ref{eq:design_FZL} was:
%In our case, after determination of the lens profile, we can rewrite Eq. \ref{eq:design_FZL} for all wavelengths:
%%%
\begin{equation}
A_{0,\lambda}I + k\lambda + n_{\lambda}IH - E_0H = optical\,path
\label{eq:op_FZL}
\end{equation}
%%%
For one wavelength and one type of discretized profile, the $optical\,path$ passing through all the $I$ points sample the emerging wavefront shape. We can then calculate the corresponding PSF by Fresnel propagation, and compare the energy in the central peak with the energy that would be confined to the central peak of a perfectly spherical wavefront. The ratio of these two quantities is defined as the {\it efficiency} of the FZL.

\begin{figure}[!htbp]
\begin{center}
\includegraphics[width=0.9\linewidth]{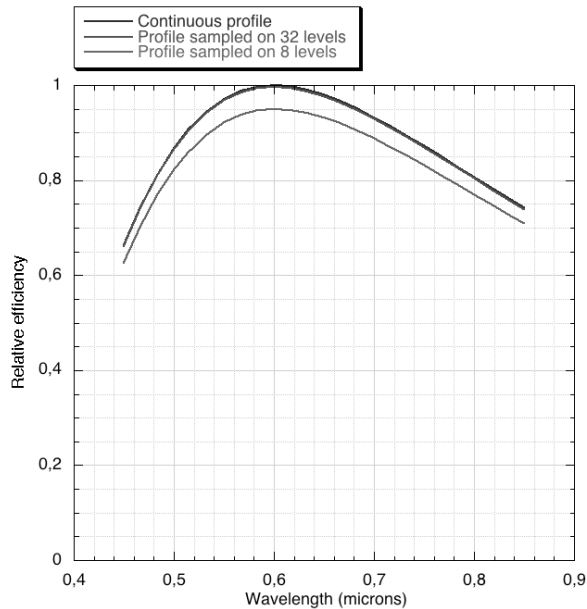}
\end{center}
\caption[FZL: evolution of efficiency]
{\footnotesize Numerically simulated efficiency as a function of wavelength for a fused silica cophased Fresnel Zone Lens, optimized for $\lambda$= 600 nm. The 3 different curves are for lenses with a continuuous profile, with a profile sampled onto 32 levels (these two efficiencies not distinguishable at this scale), and a profile sampled onto 8 levels (top to bottom curves). Our lens (Fig. \ref{fig:FZL_realized}), discretized with 128 depth levels, has an efficiency not distinguishable, on the display scale, from that of a continuous profile lens. More than 90\% efficiency is available through a $\Delta \lambda / \lambda = 0.3$ bandpass.}
\label{fig:FZL_efficiency}
\end{figure}

The computed efficiency as a function of wavelength for a cophased diverging FZL blazed at $\lambda$= 600 nm is plotted in Fig. \ref{fig:FZL_efficiency}. This is in agreement with results from Hasman (Tab. \ref{fig:table_eff}) and from Faklis \& Morris 1995 \cite{Faklis_ao_1995}.

%%%
\begin{table*}[t]
\centering
\begin{tabular}{ccc}
%\begin{tabular}{|c|c|c|}
%\hline
Profile type & Ccomputed efficiency & Theoretical limit\\
\hline
Continuous profile & 100.00\% & 100.00\%\\
Profile sampled on 128 levels & 99.98\% & 99.98\%\\
Profile sampled on 32 levels & 99.7\% & 99.7\%\\
Profile sampled on 16 levels & 98.7\% & 98.7\%\\
Profile sampled on 8 levels & 95.0\% & 95.0\%\\
Profile sampled on 4 levels & 81.2\% & 81.1\%\\
\hline
\end{tabular}
\caption {Comparison between efficiencies at 1$^{st}$ order ($\lambda_{blaze}$) computed with Eq. \ref{eq:op_FZL}%by the Fourier Transform of the wavefront state given by calculation of Eq. \ref{eq:op_FZL} for $I_{sampled}$ of different number of levels% our Fresnel propagation simulated optical train, 
, and the theoretical limits calculated by Hasman \cite{Hasman_ol_1991}.}
\label{fig:table_eff}
\end{table*}
%%%%%%%%%%
\subsection{Chromatic correction}

The targets for optical tests are pinholes, single mode optical fibers or extended sources, illuminated with narrowband or broadband spectra, $\lambda \in [400 \text{ nm} ; 950 \text{ nm}]$.

\begin{figure}[!htbp]
\begin{center}
\includegraphics[width=0.9\linewidth]{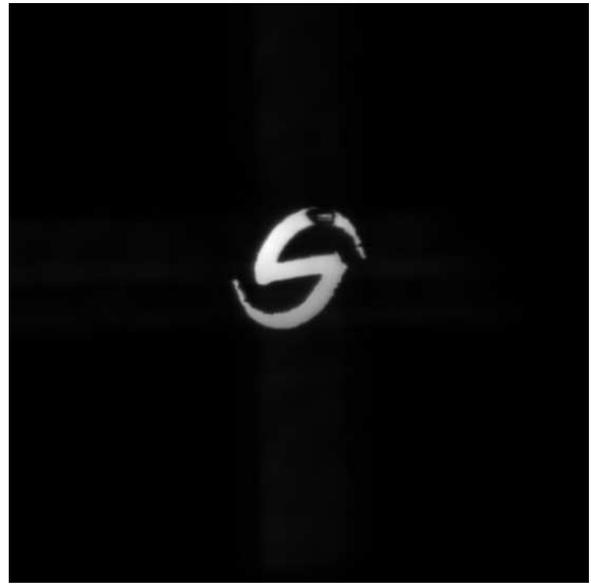}
\end{center}
\caption[Image of the 'galaxy' shaped target]
{\footnotesize A microscopic 72 arc seconds, galaxy-shaped target, laser carved into a metal foil is illuminated with a halogen source and collimated. It is then imaged by the Fresnel imager prototype. The cutting irregularities and metal bubbles that can be seen are real and not due to imaging. The faint horizontal and vertical lines are the two orthogonal diffraction spikes due to the Fresnel array.}
\label{fig:galaxy}
\end{figure}

In Fig. \ref{fig:galaxy} we show the acquired image of a galaxy-shaped target, cut out from a metal sheet and non uniformly illuminated with a halogen source. Although the clipping has been done with a UV laser machine tool, the very small linear dimensions of the target (450$\mu$m) result in a rough aspect. The product of the halogen spectrum and CCD detector sensitivity cover from 400 to 950 nm. The angular size of this "galaxy" is 72 arcsec from limb to limb, the diffraction limit of the square aperture prototype being $\lambda / C = 1.55$ arcsec at 600nm. No defocus or differential magnification can be seen, qualitatively illustrating the efficiency of the chromatic correction principle and realization.

%%%%%%%
\subsection{Angular resolution}
\begin{figure}[!htbp]
\begin{center}
\includegraphics[height=0.9\linewidth]{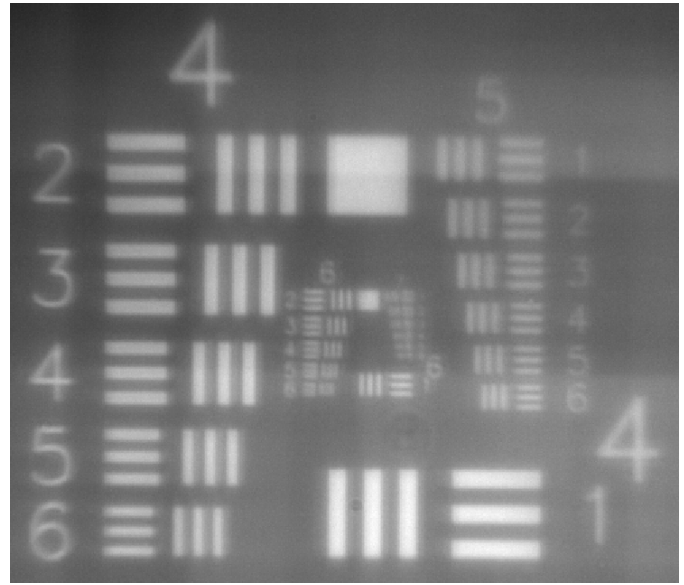}
\end{center}
\caption[Image of the USAF target]
{\footnotesize A standard  USAF test target is placed at the focal plane of the collimator, illuminated with a white led and imaged by the Fresnel imager prototype. The '6' group number associated to the '4' element number results in a number of line pairs per mm corresponding to the diffraction limit of our prototype. This image is a raw exposure, simply dark-subtracted. As this target is an extended source and is convoluted with the spikes of the PSF, high dynamic range imaging cannot be achieved in this case. However, very high dynamic range is achievable for sparse fields.}
\label{fig:mire}
\end{figure}

\begin{figure}[!htbp]
\begin{minipage}[c]{0.46\linewidth}
\begin{center}
\includegraphics[width=0.9\linewidth]{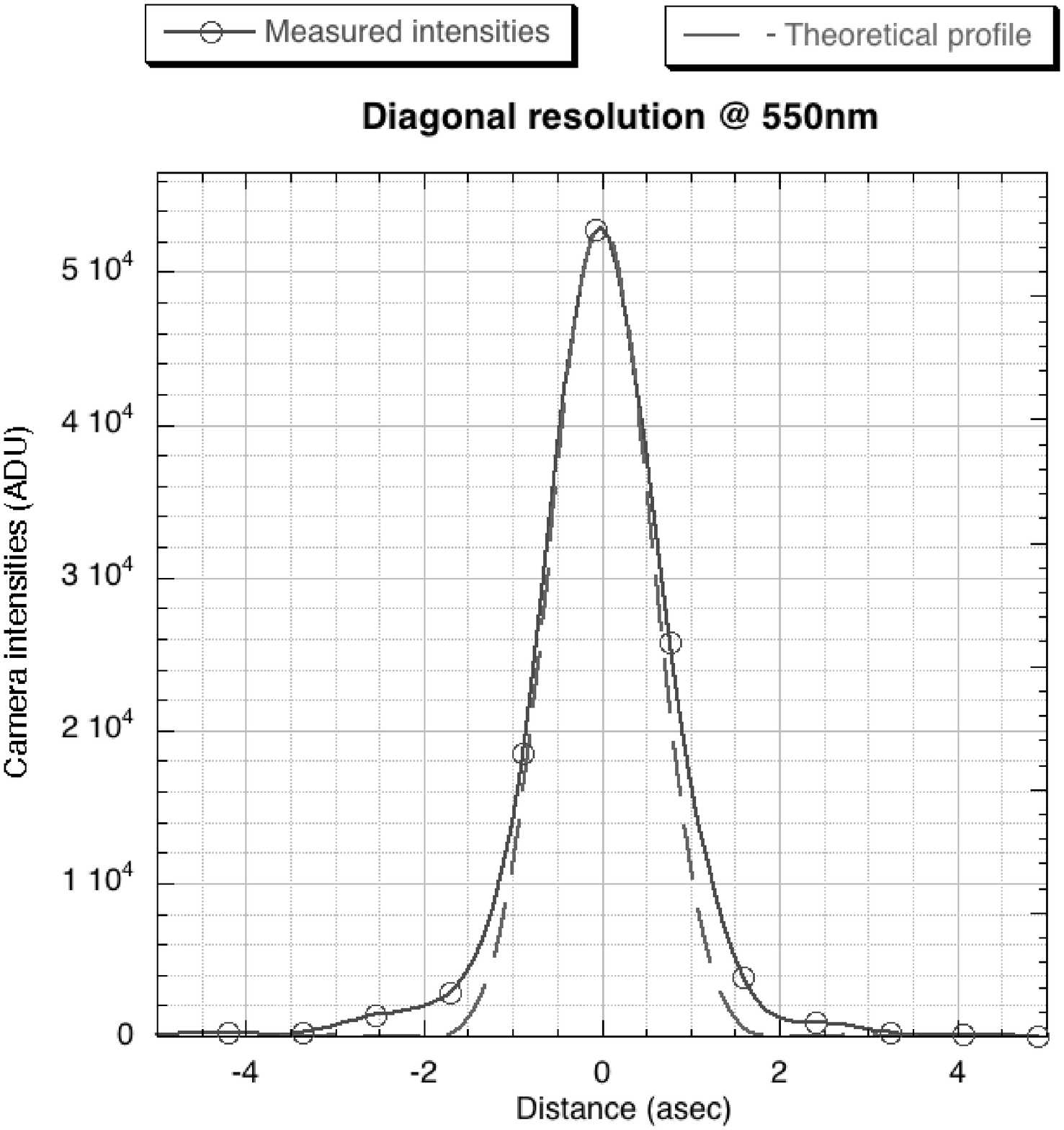}
\end{center}
\end{minipage}
\hspace*{0.5cm}
\begin{minipage}[c]{0.46\linewidth}
\begin{center}
\includegraphics[width=0.9\linewidth]{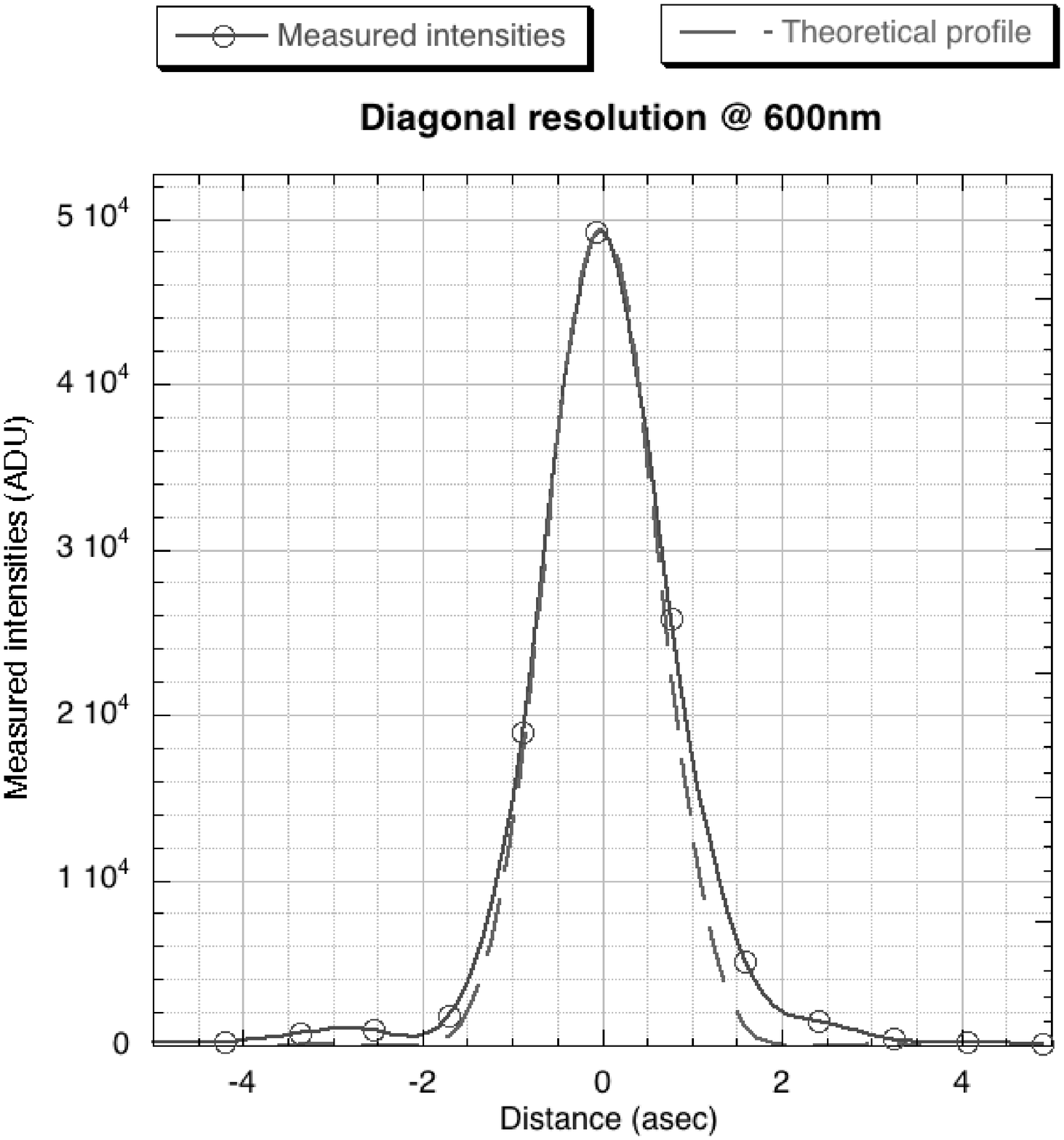}
\end{center}
\end{minipage}

\begin{minipage}[c]{0.46\linewidth}
\begin{center}
\includegraphics[width=0.9\linewidth]{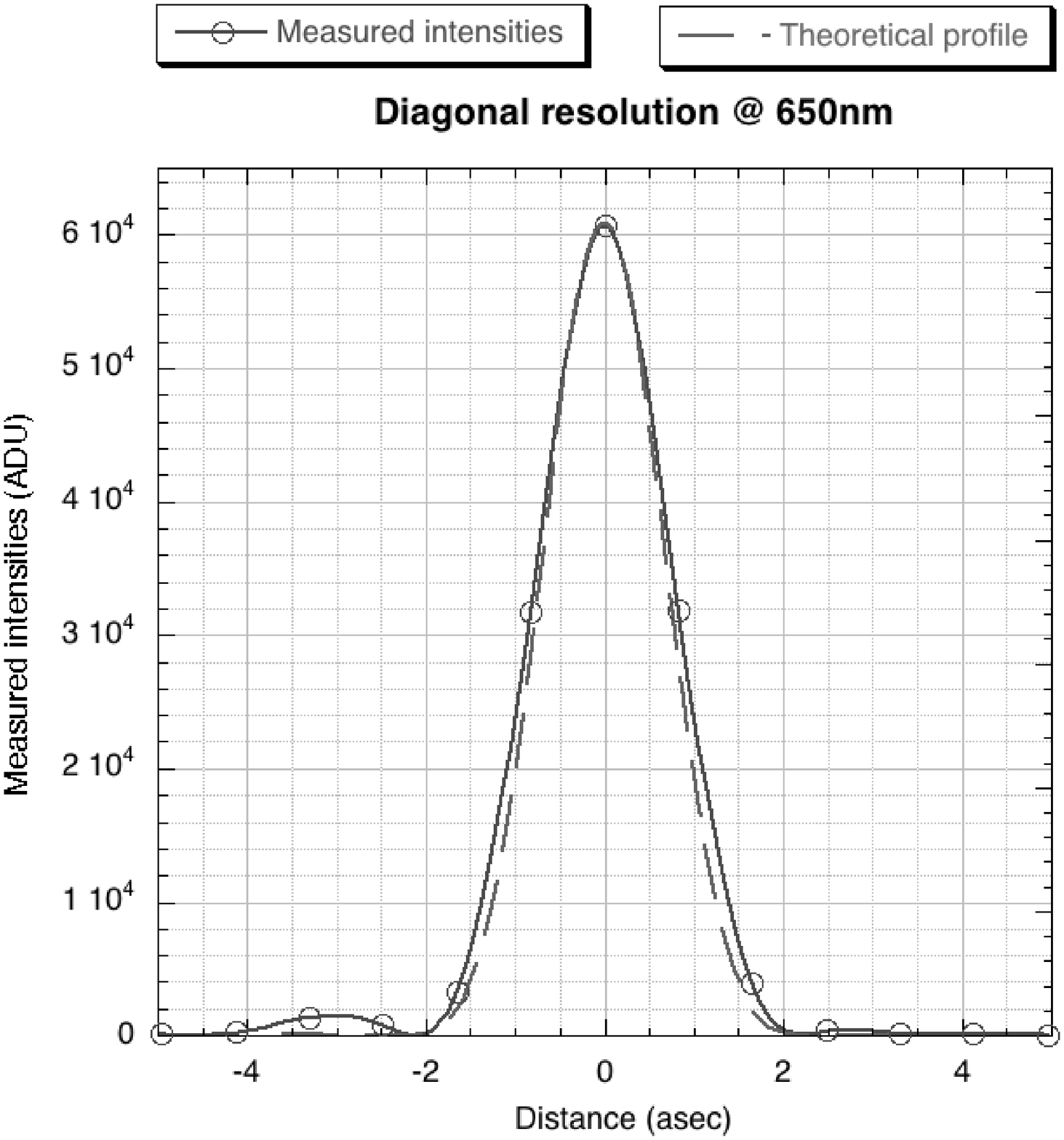}
\end{center}
\end{minipage}
\hspace*{0.5cm}
\begin{minipage}[c]{0.46\linewidth}
\begin{center}
\includegraphics[width=0.9\linewidth]{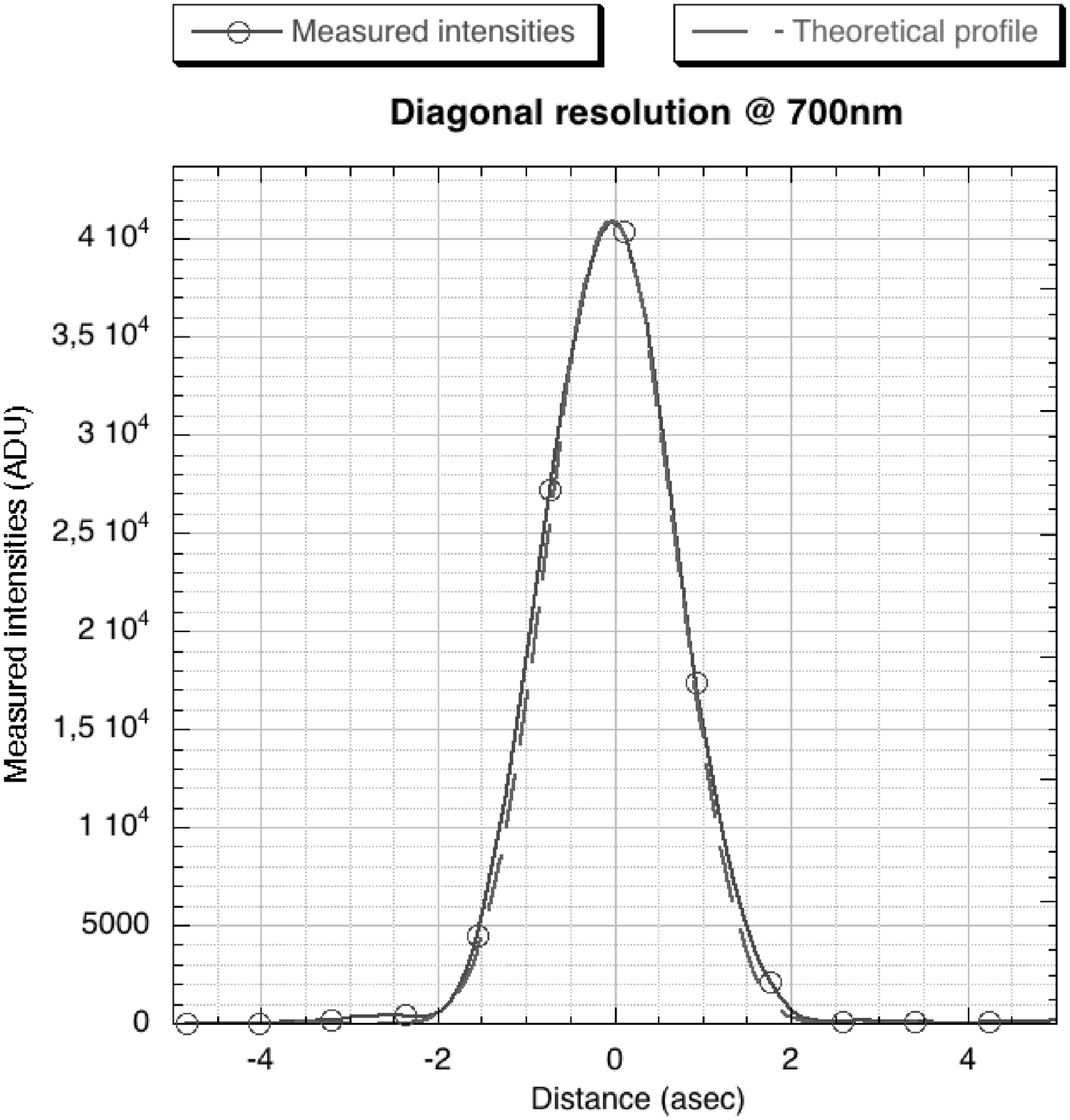}
\end{center}
\end{minipage}

\caption[Angular resolution measurements]
{\footnotesize Angular resolution measurements at four wavelengths: 550, 600, 650 and 700nm. In each graph, the theoretical profile (dashed line) is the convolution of a uniform disk source size (0.77arcsec) with a diffraction limited PSF. It is compared with the measured profile (solid line) sampled from the interpolated measurements points (circles): brightness of the camera pixels. The prototype reaches limited by diffraction for all these wavelengths, confirming the efficiency of the chromatic correction and the blazed diverging Fresnel zone lens design. Scattered light near central peak can be attributed to air turbulence observed in the clean room.}
\label{fig:angular_resolution_measures}
\end{figure}

In Fig. \ref{fig:mire} we show the acquired image of a standard USAF resolution test target illuminated with a white led. The group number '6' associated to element number '4' (number of line pairs=90.5) means a 1.52 arcsec ($\pm 0.04$) angular separation as seen from the Fresnel Imager prototype.\\
We have also measured the angular resolution of the Fresnel Imager, using a pinhole (angular diameter seen by the prototype: 0.77 arc second), illuminated with a halogen source filtered with 50nm bandwidth filters centered on 550, 600, 650 and 700 nm. The prototype is diffraction limited at all these wavelengths. The measurements are summarized in Fig. \ref{fig:angular_resolution_measures}.
%%%%%%%
%%%

%%%%%%%%%%%%%
\subsection{Flux transmission}
We have numerically simulated the image resulting from the Fresnel imager prototype by plane-to-plane Fresnel propagation. We have taken into account all its optical elements: the Fresnel array, field lens, zero order mask, blazed FZL with its manufacturing charcacteristics and the final achromatic doublet. According to these numerical simulations, our prototype yields a 4.0\% transmission when equipped with an orthogonal primary array, whereas our new multispider Soret zone plate as primary array yields a transmission of 6.3\%. 

Using the CCD, we have measured a flux ratio of 1.23 between the two types of Fresnel arrays.

%%%%%%%%%%%%%
\subsection{Dynamic range}

\begin{figure}
\begin{center}
\includegraphics[width=0.99\linewidth]{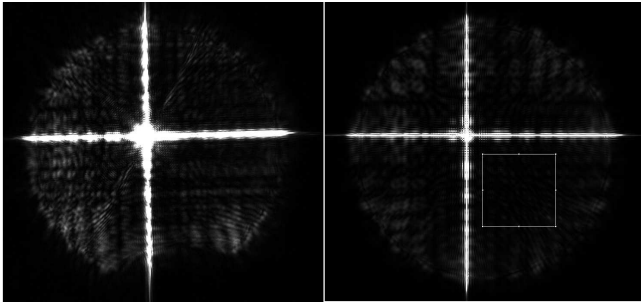}
\end{center}
\caption[Comparison of simulated and obtained PSFs]
 {\footnotesize The PSF on the left has been obtained with our prototype, using a luxeon LED with peak emission at 630 nm illuminating a monomode fiber. This image has simply been dark-subtracted. The PSF on the right is from our computer simulation, taking all the optical elements and the spectrum of the source into account. The computer simulation considers perfect optical elements except the FZL for which the commissionned jigsaw profile is used, and does not model the air turbulence. The two images are highly saturated, with the same thresholds, in order to show the faint levels of the PSF. The dynamic range is slightly better in the theoretical case than in the measured one (1 $10^{-6}$ in the square delimitation shown on the right figure, 2 $10^{-6}$ on the corresponding one on the real measurement). The slight shadow at the bottom of the left image is due to the secondary mirror of the Maksutov telescope used as field lens. This comparison validates the numerical simulation tools, which can be used to predict what can be expected with larger arrays.}
\label{fig:real_psf_vs_simulated}
\end{figure}

Fig. \ref{fig:real_psf_vs_simulated} shows the comparison between the simulated and acquired PSFs of the multispider Soret zone plate presented in Sec. \ref{sec:circ_dvpment}. We define the "dynamic range" as the ratio between the "clean" field mean level and the PSF peak maximum. With the multispider Fresnel array, the dynamic range varies from 3 $10^{-6}$ to 1 $10^{-6}$, depending on the quadrant chosen in the image produced by the prototype and its extend, while being 1 $10^{-6}$ in the numerical simulation. 

The Fresnel arrays in our prototype only have 116 Fresnel zones, are not apodized, and the zero order mask is wide compared to the resolution of the field optics. This zero order mask is placed where the interference zero order (plane wave) from the primary array is focussed by the field optics. It improves the dynamic range by blocking the uniform field illumination that would result otherwise from order zero. However, if it's too large, it causes central obstruction in the order +1 beam.  Thus, with our present prototype high dynamic range can not be achieved closer than 10 resolution elements from the central peak of the PSF. 
%We have also created a double source by the use of two optics fibers, these two sources having highly different and adjustable flux. 

With a higher number of Fresnel zones and an adapted zero order mask, numerical simulations with a non apodized square aperture and 700 Fresnel zones show that a $10^{-7}$ dynamic range is obtained over nearly the entire field, except the four thin spikes, at a five resolution elements radius from the center of the PSF peak. With a 700 zones apodized array, the dynamic range reaches $10^{-8}$ as close as two resolution elements from center in narrowband ($\frac{\Delta \lambda}{\lambda} \text{ typ. } \frac{1}{10}$), but the overall transmission is reduced by a factor of 5. We are working on a compromise, and are considering the implementation of a PIAA setup.
%%%%%

%%%%%%%%%%%%%%%%%%%%%%%%
\section{Conclusion}

At Laboratoire d'Astrophysique de Toulouse-Tarbes, we have constructed a prototype, illustrating the efficiency of our Fresnel array concept. This prototype uses a lightweight metal Fresnel array  as main aperture optics and efficiently corrects the chromaticism, using a custom built FZL. 
This results in diffraction limited images, highly contrasted on compact sources.

We have also developed a complete tool for numerical simulation and assessment of large Fresnel imagers. 

We are building a generation II prototype, featuring a 20 cm diameter, 700 zone Fresnel array, and the associated focal instrumentation module. This Gen II prototype will be placed in parallel to the 19-meter long tube of a refractor on an equatorial mount, the {\it 76 cm refractor} at Observatoire de Nice. For the next two years, we plan to test the limits of this concept on highly contrasted astrophysical targets.

A phase zero study is also under progress at Centre National d'Etudes Spatiales. It has shown that systems using thin foil Fresnel arrays up to 15-meter apertures can be built for space with ``off the shelf'' technology. It has also been shown that the guiding and navigation control tolerances are fairly within reach of present technology for a Two-Spacecraft formation flying Fresnel Imager orbiting the L2 Sun-Earth Lagrangian point. Of course, funding, and consequently validation of the scientific program is the main issue. Following the Cosmic Vision proposal (Koechlin 2008 \cite{Koechlin_expa_2008}), a working group is being set up to define the astrophysical themes that can be addressed with a 4-m to 40-m aperture space-based Fresnel Imager. Collaborations are welcome !

{\small{\it{Part of this work was funded by Centre National d'Etudes Spatial and Thales Alenia Space. The authors wish to thank the anonymous Referee for his/her remarks, and Lars E. Kristensen for re-reading the article.}}}

%\bibliographystyle{unsrt}
%\bibliography{./biblio}

\begin{thebibliography}{10}

\bibitem{Baez_nature_1960}
A.~Baez.
\newblock {A Self-supporting Metal Fresnel Zone-plate to focus Extreme
  Ultra-violet and Soft X-Rays }.
\newblock {\em Nature}, 186:958, June 1960.

\bibitem{Baez_josa_1961}
A.~Baez.
\newblock Fresnel zone plate for optical image formation using extreme
  ultraviolet and soft x radiation.
\newblock {\em Journal of the Optical Society of America}, 51(4):405--412,
  1961.

\bibitem{Chesnokov_sb_1993}
Yuri~M. Chesnokov.
\newblock A space-based very high angular resolution telescope.
\newblock {\em Space Bulletin}, 1(2):18--21, 1993.

\bibitem{Hyde_ao_1999}
Roderick~A. Hyde.
\newblock Eyeglass. 1. very large aperture diffractive telescopes.
\newblock {\em Applied Optics}, 38(19):4198--4212, 1999.

\bibitem{Massonnet_brevet_cnes}
Didier Massonnet.
\newblock Un nouveau type de t\'{e}lescope spatial - $\text{B}$revet
  $\text{CNES}$ - ref. 03.13403, 2003.

\bibitem{Koechlin_aa_2005}
Laurent Koechlin, Denis Serre, and Paul Duchon.
\newblock High resolution imaging with fresnel interferometric arrays:
  suitability for exoplanet detection.
\newblock {\em Astronomy \& Astrophysics}, 443:709--720, 2005.

\bibitem{Faklis_oe_1989}
Dean Faklis and George~Michael Morris.
\newblock Broadband imaging with holographic lenses.
\newblock {\em Optical Engineering}, 28(6):592--598, 1989.

\bibitem{Koechlin_expa_2008}
L.~{Koechlin}, D.~{Serre}, P.~{Deba}, R.~{Pell{\'o}}, C.~{Peillon},
  P.~{Duchon}, A.~I. {Gomez de Castro}, M.~{Karovska}, J.-M. {D{\'e}sert},
  D.~{Ehrenreich}, G.~{Hebrard}, A.~{Lecavelier Des Etangs}, R.~{Ferlet},
  D.~{Sing}, and A.~{Vidal-Madjar}.
\newblock {The fresnel interferometric imager}.
\newblock {\em Experimental Astronomy}, 23:379--402, March 2009.

\bibitem{Serre_spie_2007}
Denis Serre, Laurent Koechlin, and Paul Deba.
\newblock Fresnel interferometric arrays for space-based imaging: testbed
  results.
\newblock In Howard~A. MacEwen and James~B. Breckinridge, editors, {\em
  UV/Optical/IR Space Telescopes: Innovative Technologies and Concepts III, in
  Proceedings of the SPIE.}, volume 6687 of {\em Presented at the Society of
  Photo-Optical Instrumentation Engineers (SPIE) Conference}, page 66870I,
  September 2007.

\bibitem{Nisenson_apj_2001}
P.~{Nisenson} and C.~{Papaliolios}.
\newblock {Detection of Earth-like Planets Using Apodized Telescopes}.
\newblock {\em The Astrophysical Journal}, 548:L201--L205, February 2001.

\bibitem{Guyon_aa_2003}
O.~{Guyon}.
\newblock Phase-induced amplitude apodization of telescope pupils for
  extrasolar terrestrial planet imaging.
\newblock {\em Astronomy \& Astrophysics}, 404:379--387, June 2003.

\bibitem{Soret_1875}
J.~L. Soret.
\newblock Sur les ph\'{e}nom\`{e}nes de diffraction produits par les
  r\'{e}seaux circulaires.
\newblock {\em Archives des Sciences physiques et naturelles}, 52:320--337,
  1875.

\bibitem{Kipp_nature_2001}
L.~{Kipp}, M.~{Skibowski}, R.~L. {Johnson}, R.~{Berndt}, R.~{Adelung},
  S.~{Harm}, and R.~{Seemann}.
\newblock {Sharper images by focusing soft X-rays with photon sieves}.
\newblock {\em Nature}, 414:184--188, November 2001.

\bibitem{Schupmann_1899}
L~Schupmann.
\newblock Die medial-fernrohre: eine neue konstruktion f$\ddot{u}$r grosse
  astronomische instrumente.
\newblock {\em Teubner B G}, 1899.

\bibitem{Faklis_ao_1995}
Dean Faklis and George~Michael Morris.
\newblock Spectral properties of multiorder diffractive lenses.
\newblock {\em Applied Optics}, 34(14):2462--2468, 1995.

\bibitem{Swanson_oe_1989}
Garry~J Swanson and Wilfrid~B Veldkamp.
\newblock Diffractive optical elements for use in infrared systems.
\newblock {\em Optical Engineering}, 28(6):605--608, 1989.

\bibitem{Hasman_ol_1991}
E~Hasman, N~Davidson, and A~A Friesem.
\newblock Efficient multilevel phase holograms for $\text{CO}_2$ lasers.
\newblock {\em Optics Letters}, 16(6):423--425, 1991.

\bibitem{Levy_josa_2001}
Uriel Levy, Devid Mendlovic, and Emanuel Marom.
\newblock Efficiency analysis of diffractive lenses.
\newblock {\em Journal of the Optical Society of America}, 18(1):86--93, 2001.

\end{thebibliography}
% in the submitted version: must replace the two lines above with the content of proto_v1x.bbl

%%%%%
\newpage
%%%%%%

%%%%%%
%\newpage
%%%%%%

%%%%%%%
%\newpage
%%%%%

%%%%%
%\newpage
%%%%%

%%%%%%
\newpage
\vspace*{3cm}
%%%%%%

%%%%%
%\newpage
%%%%%

%%%%%%
\newpage
%%%

%%%
%\newpage
%%%

%%%%%%
%\newpage
%%%%%%

%%%%%%%%%%%%%
%\newpage
%%%%%%

%%%%%%
%\newpage
%%%%%%

%%%%%%
%\newpage
%%%%%%

%%%%%%
%\newpage
%%%%%

%\cleardoublepage
%\listoffigures

\end{document}